\documentstyle[12pt,epsf]{article}
\begin{document}


\def\gsim{\lower0.5ex\hbox{$\stackrel{>}{\sim}$}}
\def\lsim{\lower0.5ex\hbox{$\stackrel{<}{\sim}$}}
\def\eq#1{Eq.~(\ref{eq:#1})}


\def\xfigi
{
\begin{figure}[bht]
 \vspace{2cm}
 \epsfxsize=0.5\hsize
 \centerline{\epsfbox{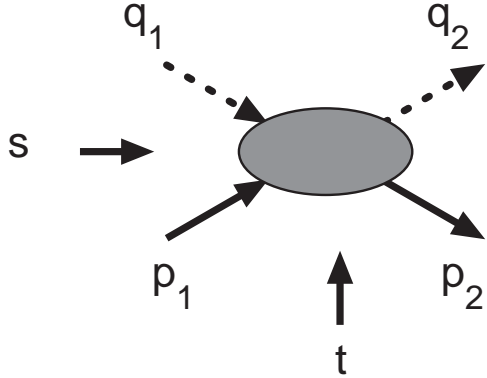}}
 \caption{
  The fundamental DVCS process:
scattering of a deeply virtual photon ($q_1$) off a nearly
on-shell light quark ($p_1$)
to produce an on-shell photon ($q_2$) and a quark ($p_2$).
 }
 \label{fig:one}
\end{figure}
}
\def\xfigii
{
\begin{figure}[bht]
 \vspace{2cm}
 \epsfxsize=\hsize
 \centerline{\epsfbox{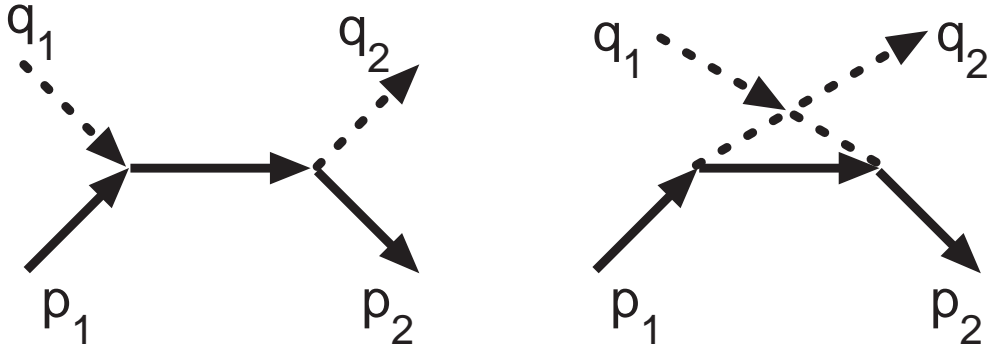}}
 \caption{   Born graphs for  the DVCS process.}
 \label{fig:two}
\end{figure}
}
\def\xfigiii
{
\begin{figure}[bht]
 \vspace{2cm}
 \epsfxsize=1.0\hsize
 \centerline{\epsfbox{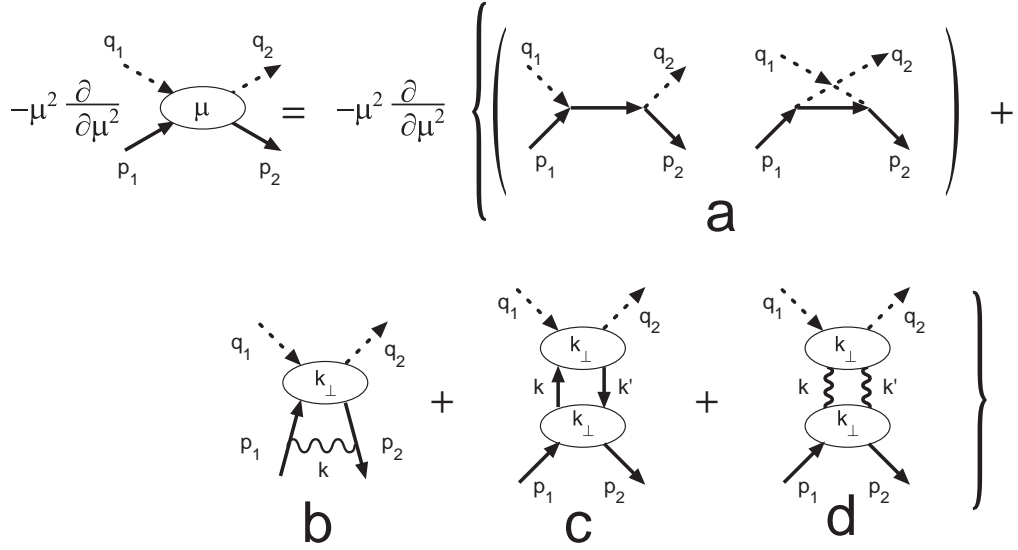}}
 \caption{
  The general form of the evolution equations for DVCS.
 }
 \label{fig:three}
\end{figure}
}
\def\xfigiv
{
\begin{figure}[bht]
 \vspace{2cm}
 \epsfxsize=0.5\hsize
 \centerline{\epsfbox{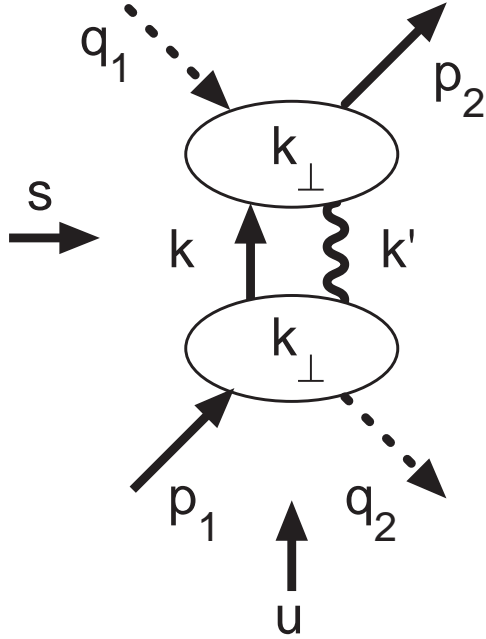}}
 \caption{
 The two-particle intermediate state for the backward DVCS process.
 }
 \label{fig:four}
\end{figure}
}
\def\xfigv
{
\begin{figure}[bht]
 \vspace{2cm}
 \epsfxsize=\hsize
 \centerline{\epsfbox{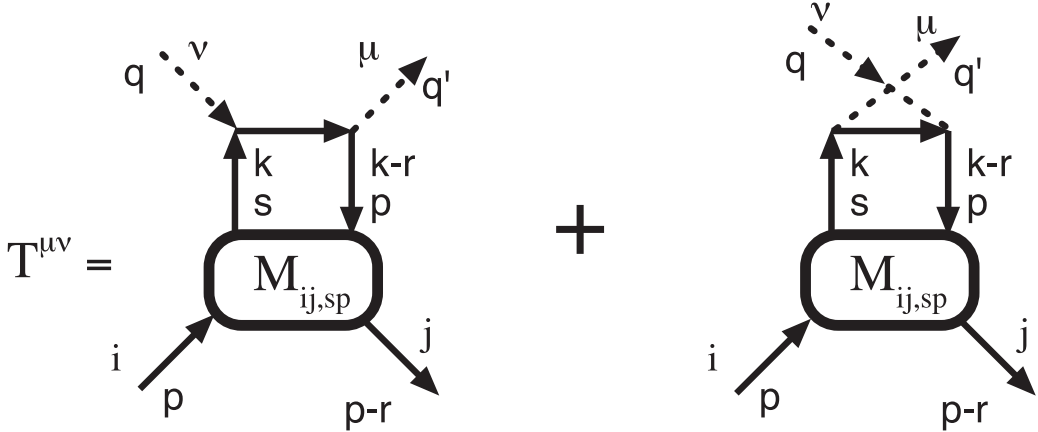}}
 \caption{
  The graphs for the non-forward structure functions
 }
 \label{fig:five}
\end{figure}
}


\begin{titlepage}

\hfill
\begin{tabular}{l}
\\
hep-ph/9812271 \\
SMU-HEP/98-08 
\end{tabular}

\vspace{2cm}

\begin{center}

\renewcommand{\thefootnote}{\fnsymbol{footnote}}
{\LARGE
Asymptotic high energy behavior of the deeply virtual
Compton scattering amplitude\footnote[2]{%
\noindent
 This work is supported in part by
 Grant INTAS-RFBR-95-311,
 Grant RFBR-98-02-17629,
 the US Department of Energy, and the
Lightner-Sams Foundation. }
}
\renewcommand{\thefootnote}{\arabic{footnote}}

\vspace{1.25cm}

{\large
B.I.~Ermolaev
\\
A.F.~Ioffe Physico-Technical Institute, St.~Petersburg, 194021, Russia
\\
 F.~Olness
\\
Southern Methodist University, Dallas, TX 75275, USA
\\
 A.G.~Shuvaev
\\
 St.~Petersburg Nuclear Physics Institute, Gatchina, St.~Petersburg, 188350,
Russia }

\vspace{1.25cm}

\end{center}

\vfil

\begin{abstract}
 We compute  the  deeply virtual Compton scattering  (DVCS) amplitude for
forward and backward scattering in the asymptotic limit.
 Since this calculation does not assume ordering of the transverse
momenta, it includes important logarithmic contributions that are
beyond those summed by the DGLAP evolution.
 These contributions lead to a power-like behavior for the forward DVCS
amplitude.
 \end{abstract}

\vfil
\end{titlepage}
\newpage
\section{Introduction}

The deeply virtual Compton scattering (DVCS) has received
attention recently since it provides a new and interesting
theoretical framework for perturbative and nonperturbative
QCD applications.\cite{Ji,Rad,DVCS}
 The basic mechanism of this process is
reminiscent of the usual deeply inelastic (DIS) lepton-hadron
scattering process which  begins with the absorption of a
virtual photon  by the quark constitutient.  The DVCS differs
from DIS in the second step when the same  quark
constitutient then emits an {\it on-shell} photon.
 The kinematics of the DVCS  reaction are characterized by a
high invariant energy and a large virtuality of the {\it
initial} photon compared to the quark mass.

The DVCS amplitude is, in a sense, a generalization of  the
conventional deep inelastic structure (DIS) functions, and
can be expressed using asymmetric distribution functions.
 To incorporate the radiative corrections into the DVCS amplitude,  the
conventional DGLAP evolution equations\cite{DGLAP} have been used
\cite{Ji,Rad,DVCS} (for review on the DGLAP evolution for the DVCS see
\cite{review} and references therein). As a typical for the DGLAP result,  the
exponential asymptotical high energy behavior of the DVCS amplitude was
obtained. Such asymptotics are true for the kinematical region
where the invariant total energy and the virtuality of the off-shell
photon are of the same order, i.e. in so-called the "hard" kinematics.
However, in order to obtain the DVCS asymptotics in the other kinematical
regions, and in particular in the "semi-hard"(or  the Regge type) one,
with the invariant
total energy being much greater than the off-shell photon virtuality,
a generalized set of evolution equations is required.

Although predictions based on the DGLAP evolution yield good
agreement with present experimental deep inelastic
scattering data,\cite{A,BV,BGR}
from the theoretical point of view this approach is  insufficient when
the
kinematics are of the Regge type where the total invariant
energy is much greater than the other invariants, including
the virtuality of the deeply virtual photon.  In this
kinematic region, the DGLAP equations can not account for
contributions independent of this virtuality.
 Taking these contributions into  account for the DIS
structure functions led to a power-like dependence on the
total energy both for photon scattering from an unpolarized
quark (the BFKL Pomeron\cite{BFKL}) and for the scattering
from a polarized quark.\cite{BER}

\xfigi  

Naturally, one can expect a similar change in the asymptotic
behavior of the DVCS amplitude in the Regge limit.  To
demonstrate this behavior, let us consider the scattering of
a deeply virtual photon off a nearly on-shell light quark
which produces a photon and a quark---both on-shell, ({\it cf.},
Fig.1).
 In Fig.1, $s$ , $t$ and $u$  are the standard Mandelstam variables:
 \begin{eqnarray}
s &=& (p_1 + q_1)^2  \nonumber \\
t &=& (q_2- q_1)^2   \\
u &=& (q_2 - p_1)^2 \quad . \nonumber
\label{eq:one}
 \end{eqnarray}

\xfigii  

 In the Born approximation,
only the two Feynman graphs in Fig.2 contribute to the
DVCS process:
 Beyond the Born approximation the value of the
QCD-radiative corrections strongly depend on the
kinematics of the scattering. In particular,
 in the two  regions
 \begin{eqnarray}
s &\sim& - u \gg -t \qquad  \qquad   {\rm Forward  \ Region}   \label{eq:for}
\\[10pt]
s &\sim& - t \gg -u \qquad  \qquad   {\rm Backward \ Region}   \label{eq:bac}
\label{eq:two}
 \end{eqnarray}
 the QCD radiative corrections are large,  whereas in the region
 $s \sim -t \sim -u$ they are comparatively small.
In the center of mass system (CMS),
 $s \sim - u \gg -t$ corresponds to the {\it Forward Region} (\eq{for})
 (small scattering angles) , and
 $s \sim - t \gg -u$ corresponds to the {\it Backward Region} (\eq{bac}).

  Both of the above regions
are of the Regge type. When
\begin{eqnarray}
s \gg Q^2,
\label{eq:regge}
\end{eqnarray}
where $Q^2 = -q_1^2$ is the photon virtuality, the Regge theory predicts a
power-like
 $s$-dependence for the DVCS amplitude in each of the regions.
 Such behavior is result of summing the QCD-radiative corrections to all orders
in the QCD-coupling.

 Below we obtain the asymptotical behavior of the DVCS amplitudes for the
"semi-hard" kinematical regions (\eq{for},\eq{regge}; \eq{bac},\eq{regge}) in
the leading
logarithmic
approximation (LLA)  by solving  the
Infrared Evolution Equations (IREE) \cite{KL,EL} for them. 
 As in the case of the DGLAP evolution, in order to obtain the evolution in the Regge
regions above, we consider the DVCS from an on-shell quark.
 Recently, the IREE method has been used to obtain the
small-$x$ asymptotic limit of the DIS
spin-dependent structure functions \cite{BER} .
While references  \cite{Ji,Rad,DVCS} include consideration of the hadronic 
component, 
we study only the partonic
part of the DVCS which can be investigated with the perturbative methods
of QCD.  Since we consider the DVCS from a quark,  our results
depend on the factorization scale $\mu$; this factorization scale dependence
will cancel
out in the expression for the DVCS from a hadron. 
Operationally, within the LLA-accuracy, this 
effects the replacement of $\mu$ by some mass factor of order of the hadron 
mass.  

The paper is organized
as follows: In Sect.~2 we briefly review the IREE method we shall
use to obtain the
expression for the DVCS amplitude in the forward region, \eq{for}, \eq{regge}.
In Sect.~3 we investigate the DVCS amplitude in the backward region,
\eq{bac}, \eq{regge}.
In Sect.~4 we calculate the non-forward structure functions for DVCS in the
Regge region \eq{regge}, and express them in terms of the Reggeons
corresponding to on-shell scattering.
Finally, in Sect.~5 we discuss the implications of these results.

\section{Forward DVCS}

Our calculation will follow the general prescription of the IREE
method given in Ref.~\cite{KL,EL}.
 All the virtual momenta, $k_i$, can be expressed through
the Sudakov parameterization as follows:
 \begin{eqnarray}
k_i          &=& \alpha_i p_1 + \beta_i q^{\prime}_1  + k_{\perp} \quad , \\
q^{\prime}_1 &=& q_1 - {q_1^2 \over 2p_1 \cdot q_1} \ p_1 \quad , \\
k_{\perp} \cdot p_1 &=& k_{\perp} \cdot q_1 = 0 \quad ,
\label{eq:four}
 \end{eqnarray}
where we have neglected the masses of incoming and outgoing quarks.

In order to avoid infrared singularities, it is necessary to
introduce an infrared cut-off $\mu$ in transverse momentum space so that
 \begin{eqnarray}
k_i^{\perp} &>& \mu \quad .
\label{eq:five}
 \end{eqnarray}
 In the LLA, such regularization does not violate gauge invariance;
therefore the Gribov bremsstrahlung theorem\cite{G} is applicable
in this case. ({\it Cf.},  Ref.~\cite{EFL,CE}.)
 The essence of this theorem is that the virtual gluon with the minimal
$k_{\perp}$ can be factorized out of the amplitude. Consequently, the principle
contribution comes from graphs where the gluon propagator is attached
{\it only} to external lines. Additionally,  the  integrations
over the remaining gluons do not depend on the {\it longitudinal}
momentum of this factorized gluon.
 There is a dependence on the factorized gluon's {\it transverse}
momentum---this  $k_{\perp}$ becomes a new infrared cut-off for the
remaining integrations.

Since QCD-radiative corrections in the LLA do not violate
the spin structure of the
Born approximation, the forward DVCS amplitude, $M_F$, is given by:
 \begin{eqnarray}
M_F &=& M_F^{Born}
\ \  \Phi \left[ \ln(s/\mu^2), \ln(Q^2/\mu^2), \ln(-t/\mu^2) \right]
\equiv
M_F^{Born} \ \  \Phi \left[ z_1, z_2, z_3 \right]
\quad  ,
\label{eq:six}
 \end{eqnarray}
where we define
 \begin{eqnarray}
z_1 &=&  \ln(s/\mu^2)   \quad , \nonumber \\
z_2 &=&  \ln(Q^2/\mu^2) \quad , \nonumber \\
z_3 &=&  \ln(-t/\mu^2)  \quad .
\label{eq:eight}
 \end{eqnarray}

 Although  the value of $\mu^2$ is not fixed,  we presume  it
is much less than any of $s$, $Q^2$, or $-t$.
Instead of setting $\mu^2$ to a fixed value, let us differentiate
$M_F$ with respect to $\mu$ to obtain our IREE:
 \begin{eqnarray}
-\mu^2 \
  \frac{\partial M_F}{\partial \mu^2}
&=&
  \frac{\partial M_F}{\partial z_1}
+ \frac{\partial M_F}{\partial z_2}
+ \frac{\partial M_F}{\partial z_3}
\qquad .
\label{eq:ireelhs}
 \end{eqnarray}
 The LHS of \eq{ireelhs} provides half of the IREE
we desire for $M_F$, ({\it cf.},  Fig.3).
In order to obtain the remaining half of the IREE,
we should look for the virtual quark or
a gluon with the minimal transverse momentum.

\xfigiii  

If the particle with minimal transverse momentum
is a gluon, then it can be  factorized,
(graph (b) in RHS of Fig.3).
Integration over the longitudinal momentum of the gluon
yields (after differentiation over $\ln(\mu^2))$,
 \begin{eqnarray}
&& -2\lambda z_3 M_F
\label{eq:nine}
 \end{eqnarray}
with
 \begin{eqnarray}
\lambda &=& \alpha_s C_F/4\pi \quad ,
\label{eq:ninea}
 \end{eqnarray}
$C_F = (N^2-1)/2N$, and $N=3$.

If the particle with minimal transverse momentum
is a quark, then the process is more involved.
  In order to evaluate the leading-log (LL) contribution,
there should be another
quark on the ladder with minimal $k_{\perp}$ from the same ladder rung
(graph (c) in RHS of Fig.3).
However, as long as we keep $-t \gg \mu^2$, the LL
contribution arises only when $k^2_{\perp} > -t$.
Then the integration region
over $k_{\perp}$ does not include $\mu$ as the lower limit when
$-t\gg  \mu^2$, and the whole term (c) in Fig.3 will be independent of
$\mu$, and any derivatives w.r.t. $\mu$ will vanish.

By the same reasoning, we
can dismiss the case where, instead of quarks, two ladder gluons  have
minimal values of $k_{\perp}$, (corresponding to graph (d) in Fig.3).

The Born contribution (graph (a) in Fig.3) also does not depend on $\mu$
when $-t\gg \mu^2$, and this term also vanishes when
differentiated w.r.t. $\mu$.

We therefore find the RHS of the IREE is given solely by graph (b) in Fig.3,
and we obtain
 \begin{eqnarray}
&& \frac{\partial M_F}{\partial z_1} + \frac{\partial M_F}{\partial z_2}
+ \frac{\partial M_F}{\partial z_3} = -2\lambda z_3 M_F
\label{eq:ten}
 \end{eqnarray}
The solution of this IREE is:
 \begin{eqnarray}
M_F &=& \phi (z_1 - z_3, z_2 - z_3) \ e^{-\lambda z_3^2}
\label{eq:eleven}
 \end{eqnarray}
with $\phi$ to be defined below in \eq{fourteen}.

We will discuss now the kinematic region
where
 \begin{eqnarray}
s &>& Q^2 > -t \quad .
\label{eq:twelve}
 \end{eqnarray}
We then have $-t = \mu^2$,  which implies
$z_3 = \ln(-t/\mu^2) = 0$, and consequently
$\phi (z_1 - z_3, z_2 - z_3) \to \phi (z_1 , z_2)$.
In this kinematic limit we define
 \begin{eqnarray}
\widetilde{M_F} &=&  \phi(z_1, z_2)
\label{eq:fourteen}
 \end{eqnarray}
where  $\widetilde{M_F}$ is the DVCS amplitude for the scattering at
(nearly) zero angle, {\it i.e.} for $-t=\mu^2$.
 We define the function  $\phi$ using the  matching condition  of
\eq{fourteen}; therefore, we require an independent calculation of
$\widetilde{M_F}$ to obtain $\phi$.
 This amplitude does not depend on $z_3$.
 Thus, we have re-expressed the scattering amplitude in terms of
a simpler set of variables,
and eliminated one variable (namely, the smallest one).
Let us note that when $t \neq 0$, the
DVCS, even when regarded as the elastic $2 \to 2$ - process, is always
accompanied by bremsstrahlung consisting of photons and gluons with
energies less than $\sqrt{-t}$. A part of them have energies less than
$\mu$. Such very soft radiation does not affect kinematics of the DVCS. In
the LLA it would factorize out of the
scattering
amplitude  (for details see \cite{EFL}),  and would cancel with virtual
gluon/photon contributions of the
same energy scale in expressions for the cross sections. Through the
present paper we neglect both the emission and the radiative corrections
for bremsstrahlung with energies below $\mu$. So, within LLA-accuracy
we identify $\tilde{M_F}$ as the usual DVCS-amplitude with $t = 0$.
While we begin our calculation with a complicated process involving off-shell
exchanges, our final result involves
 elastic scattering amplitudes with only on-shell
particles. For the DVCS process, the relevant process is the
elastic quark-photon scattering amplitude with all the particles on-shell.

We now construct the IREE for $\widetilde{M_F}$ as defined in \eq{fourteen}.
Recall that we are working in the limit where $z_3 = \ln(-t/\mu^2) = 0$.
 In the following, we will show that
the only contributions to $\widetilde{M_F}$ come from graphs
(a), (c), (d) of Fig.3.
Let us now demonstrate this result.
 When the incoming and the outgoing quarks in Fig.1 are
polarized, there are spin-dependent, as well as spin-independent,
contributions to the DVCS amplitude. In
the Born approximation, the spin-independent contribution is symmetric
with respect to permutations of $s$ and $u$, whereas the spin-dependent
contribution is antisymmetric. This is also true when radiative
corrections in the LLA are taken into account.
Therefore,
the spin-independent part of the DVCS amplitude has a positive signature,
the spin-dependent   part  has a negative signature.

Consequently, we find it convenient to decompose the DVCS amplitude
into a spin-dependent amplitude $N_F$, and a spin-independent amplitude $U_F$:
 \begin{eqnarray}
M_F &=& N_F + U_F
\label{eq:fifteen}
 \end{eqnarray}
We make the same decomposition for the $\widetilde{M_F}$,
the DVCS amplitude
in the $z_3=0$ limit:
 \begin{eqnarray}
\widetilde{M_F} &=&  \widetilde{N_F} +  \widetilde{U_F}
\label{eq:fifteena}
 \end{eqnarray}
 The $t$-dependence of $M_F$
is given by \eq{eleven} provided $\widetilde{M_F}$ is known.

Let us first discuss the spin-dependent part,
$\widetilde{N_F}$, of the
DVCS amplitude $\widetilde{M_F}$.
If we neglect mass of the initial/final state
quark, the quark spin is then collinear to its momentum.
Since the transverse spin
contribution is proportional to the quark mass,
it is negligible in this limit.

For simplicity, we
consider the case where the QCD-radiative corrections to the Born
approximation correspond to adding extra gluon propagators to both
graphs of Fig.2.
 In this case,  term (d) of Fig.3  (with a two-gluon intermediate state)
does not contribute to to the RHS of IREE.
Neither terms (a) and (b) also do not contribute because $t=0$.

We implicitly define $\widetilde{F}$ via a Mellin transform
 \begin{eqnarray}
\widetilde{N}_F(z_1, z_2) &=& \int_{-\imath\infty}^{\imath\infty} \
\frac{d\omega}{2\pi\imath} \ e^{z_1\omega} \ \xi(\omega)  \
\widetilde{F}(\omega, z_2)
\label{eq:seventeen}
 \end{eqnarray}
with the negative signature factor
 \begin{eqnarray}
\xi(\omega) &=&  \frac{e^{-\imath\pi\omega} - 1}{2} \approx
\frac{-\imath\pi\omega}{2} \quad ,
\label{eq:eighteen}
 \end{eqnarray}
 The IREE for the spin-dependent part takes the following form:
 \begin{eqnarray}
\left(\frac{\partial}{\partial z_2}  + \omega\right) \widetilde{F}(\omega, z_2)
&=&
\frac{1}{8\pi^2} \widetilde{F}(\omega, z_2) f_0(\omega)
\label{eq:nineteen}
 \end{eqnarray}
where
$f_0$ is the Mellin amplitude for quark-quark elastic scattering  with
all quarks on-shell.
 It corresponds to the lowest blob of graph (c) in Fig.3., and
explicit expression  was obtained  in Ref.~\cite{KL}:
 \begin{eqnarray}
f_0 &=& 4\pi^2
\left\{ \omega - \left[\omega^2 -\frac{g^2C_F}{2\pi^2}
\left(1 -\frac{1}{2\pi^2\omega} \psi(\omega)\right) \right]^{1/2}
\right\}  \quad ,
\label{eq:twenty}
 \end{eqnarray}
with
\begin{eqnarray}
\psi(\omega) &=& g^2 N\ \frac{d}{dy} \ln
\left(e^{y^2/4}D_{\frac{-1}{2N^2}}(y)\right) \quad ,
\label{eq:twentyb}
\end{eqnarray}
and
\begin{eqnarray}
y&=& \omega/\sqrt{g^2N/8\pi^2} \quad,
\label{eq:twentyc}
 \end{eqnarray}
where $g = \sqrt{4\pi\alpha_s}$ is the QCD-coupling (which is fixed in the LLA),
and $D_{\frac{-1}{2N^2}}$ is the parabolic cylinder (or Webber) function.

The solution of \eq{nineteen} is
 \begin{eqnarray}
\widetilde{F}(\omega, z_2) &=&
\widetilde{\phi}(\omega) \
e^{-(\omega - f_0/8\pi^2)z_2}
\quad .
\label{eq:twentyone}
\end{eqnarray}
 Thus,
 \begin{eqnarray}
\widetilde{N_F}(z_1, z_2) &=&
\int_{-\imath\infty}^{\imath\infty} \
\frac{d\omega}{2\pi\imath} \
\phi(\omega) \ \xi(\omega)
\left( \frac{s}{Q^2} \right)^{\omega} \
\left( \frac{Q^2}{\mu^2} \right)^{f_0(\omega)/8\pi^2}
\label{eq:twentytwo}
 \end{eqnarray}
 \eq{twentytwo} fixes the anomalous dimension
 \begin{eqnarray}
\gamma &\equiv& f_0(\omega)/8\pi^2
\label{eq:twentythree}
 \end{eqnarray}
for $\widetilde{N_F}$.
 The unknown function $\widetilde{\phi}(\omega)$
can be fixed by the matching condition:
 \begin{eqnarray}
\widetilde{N_F} &=& C(z_1)
\label{eq:twentyfour}
 \end{eqnarray}
when $z_2 = 0$.
 The new amplitude $C(z_1)$ in \eq{twentytwo}
is the amplitude of the elastic
Compton scattering with both photons on-shell.
The IREE for $C(z_1)$ is obtained by canceling the
$z_2$-dependence in the LHS, and
adding to RHS the contribution of term (a)  in Fig.~3
which now is $\mu$-dependent.
 Expressions for the anomalous dimensions of the
spin-dependent structure function $g_1$ incorporating
all nonleading order contributions were obtained in
Ref.~\cite{EL} in the double logarithmic
approximation.  
 The resummation of the 
${\cal O}(\alpha_s^{n+1} \, \ln^{2n}x)$ term in the 
singlet evolution equation was  performed in
Ref.~\cite{BV}.

Combining the above formulas
\eq{eleven}, \eq{fourteen}, \eq{fifteen}-\eq{twentyc}, we arrive at
the result for the  spin-dependent part of the DVCS amplitude:
 \begin{eqnarray}
N_F &=&
\frac{e^2}{g^2\ C_F } \
\int_{-\imath \infty}^{\imath \infty} \ \frac{d\omega}{2\pi\imath}
\left( \frac{s}{Q^2} \right)^{\omega} \
\frac{1}{\omega - f_0/8\pi^2} \
\left( \frac{Q^2}{\mu^2} \right)^{f_0/8\pi^2} \
e^{-\frac{\alpha_s C_F}{4\pi}\ln^2(-t/\mu^2)} \
\nonumber \\
\label{eq:twentyfive}
 \end{eqnarray}
where $e^2 = \sqrt{4\pi\alpha}$.
 To identify the contribution to the spin-dependent DVCS amplitude
from the Born process (term (a) in Fig.3), we
neglect the difference between  the quark mass and $\mu$:
 \begin{eqnarray}
N_F^{Born} &=&
\int_{-\imath\infty}^{\imath\infty} \ \frac{d\omega}{2\pi i} \
\left( \frac{s}{\mu^2}  \right)^{\omega}  \
\frac{e^2}{\omega} \
\xi(\omega) \
\label{eq:twentysix}
 \end{eqnarray}

The asymptotic high energy behavior in the forward region, \eq{for},
for $N_F$ is ({\it cf.}, Ref.~\cite{BER}):

 \begin{eqnarray}
N_F &\simeq&
\left( \frac{s}{Q^2} \right)^a \
\left( \frac{Q^2}{\mu^2} \right)^{a/2} \
e^{-\frac{\alpha_s}{4\pi}\ln^2(-t/\mu^2)}
\label{eq:twentyseven}
 \end{eqnarray}
with
 \begin{eqnarray}
a &=&
\sqrt{\frac{2\alpha_s C_F}{\pi}
\left(1 + \frac{1}{2N^2} \right)
}
\quad .
\label{eq:twentyeight}
 \end{eqnarray}

We can incorporate term (d) of Fig.3 into the IREE for $F_N$  as was done in
in Ref.~\cite{BER}.
The result is to
change $a$ in \eq{twentyeight} to
 \begin{eqnarray}
a &\simeq&
3.5  \sqrt{\frac{2\alpha_s}{\pi}\
\frac{N}{4} }
\label{eq:twentynine}
 \end{eqnarray}

We now discuss $\widetilde{U_F}$, the spin-independent part of
$\widetilde{M_F}$. Thus, we are interested in the portion of the
amplitude with the positive signature.
For $t=0$, term (b) in Fig.3 does not contribute.
In contrast to the spin-dependent part of the DVCS amplitude,
the contribution of term (c) is now small compared to term (d)
by a power of $s$; therefore we neglect term (c).
 The negligible contribution of term (c) can be anticipated
since the Regge theory predicts the asymptotic behavior
of the scattering amplitude to be $\sim s^{(j_1+j_2+..) - n+1}$,
where $n$ is the number of $t$-channel intermediate particles exchanged,
and $j_i$ are their spins.

Repeating the spin-independent calculation as in the previous
spin-dependent case, we obtain the
asymptotic behavior of $U_F$ in the Regge limit of the
forward region, \eq{for}:
 \begin{eqnarray}
U_F &\sim& R(s,Q^2) \ e^{-\frac{\alpha_sN}{4\pi}\ln^2(-t/\mu^2)}
\label{eq:thirty}
 \end{eqnarray}
where $R(s,Q^2)$ is the amplitude for Compton scattering.
Since this process is mediated by the Pomeron, we obtain
$$
R(s,Q^2) \sim
\left( \frac{s}{Q^2} \right)^{1 + \Delta_P} \
\left( \frac{Q^2}{\mu_2} \right)^{\gamma_P}
\quad,
$$
where $\Delta_P$ is the Pomeron intercept, and $\gamma_P$ is the Pomeron
anomalous dimension.

Writing the virtual photon momentum as $q_1=q^\prime -\zeta p$
DVCS amplitude can be expressed in terms of the  asymmetry
parameter $\zeta$, $Q^2=\zeta s$.

\section{Backward  DVCS}

The backward DVCS process is straightforward to analyze given the previous
forward DVCS calculation.
 Since, in the Born approximation,
the contribution of graph (b) is much greater than
graph (a), we can neglect graph (a).
 Furthermore, neither graph (c) nor graph (d) contributes to the
IREE in the backward region.

\xfigiv  

Recall that  the kinematic configuration in the backward region is
specified by (\eq{bac}):
 \begin{eqnarray}
s &\sim& - t \gg -u \qquad  \qquad   {\rm Backward \ Region}
 \end{eqnarray}
Therefore,  intermediate partons with  minimal value of
$k_{\perp}$ can  only exist in the $u$-channel (when $u \simeq 0$).
The $u$-channel two particle intermediate state can not
consist of two quarks or two gluons due to fermion charge conservation.
Instead,
a new contribution given by graph in Fig.4 can now contribute. However, it
is known that this contribution is beyond the LLA. Therefore, only terms (a)
and (b) of Fig.3 contribute to the RHS of the IREE in the backward region.

Finally, we
note that there is no difference between asymptotic behavior of the
polarized and unpolarized parts of the backward DVCS amplitude $M_B$.
$M_B$ depends on the variables $z_1$,  $z_2$, and a new variable
$z_4$ defined as:
 \begin{equation}
z_4 = \ln(-u/\mu^2) \quad .
\label{eq:thirtytwo}
 \end{equation}
For  backward scattering $z_1 \approx z_3$,
so these variables are not independent.
 In a similar manner,
there was no $z_4$-dependence for the forward DVCS amplitude because
$z_4 \approx z_1$ in this region.
 The IREE for $M_B$ is ({\it cf.}, \eq{ten}):
 \begin{equation}
\frac{\partial M_B}{\partial z_1} + \frac{\partial M_B}{\partial z_2} +
\frac{\partial M_B}{\partial z_4}  = - 2 \lambda z_1
\quad .
\label{eq:thirtythree}
 \end{equation}
with $\lambda$ defined in \eq{ninea}.
The solution for $M_B$ is
 \begin{equation}
M_B = M_B^{Born} \
\Phi_B (z_1 - z_4, z_2 - z_4) \
e^{- \lambda \ln^2 z_1}
\quad .
\label{eq:thirtyfour}
 \end{equation}

If we assume  the hierarchy $ s> Q^2 > u$,
we can fix the  unknown function $\Phi_B$ with the matching condition:
 \begin{equation}
M_B = \widetilde{M_B} \qquad {\rm for }\ z_4 =0 \quad ,
\label{eq:thirtyfive}
 \end{equation}
where $\widetilde{M_B}$ is the DVCS amplitude for the backward scattering
of a collinear incoming quark and outgoing gluon.   It satisfies
 \begin{equation}
\frac{\partial \widetilde{M_B}}{\partial z_1} +
\frac{\partial \widetilde{M_B}}{\partial z_2} = -2\lambda z_1 \widetilde{M_B}
\label{eq:thirtysix}
 \end{equation}
which yields the solution
 \begin{equation}
\widetilde{M_B} = \widetilde{\Phi}_B (z_1 -z_2) \ e^{-\lambda \ln^2 z_1}
\label{eq:thirtyseven}
 \end{equation}

Finally, we can specify $\widetilde{\Phi}_B$.
When $z_2 = 0$ ($Q^2=\mu^2$), 
$\widetilde{M_B}$ must coincide with the amplitude $C_B$
for the backward Compton scattering.
$C_B$ was calculated in  the LLA
by direct calculation
of the relevant Feynman graphs.\cite{GGF}
The IREE for $C_B$  is:
 \begin{equation}
\frac{\partial C_B}{\partial z_1} = -2\lambda z_1 C_B
\label{eq:thirtyeight}
 \end{equation}
with the boundary condition
 \begin{equation}
C_B = C_B^{Born}  \qquad {\rm for }\ z_1 =0 \quad (s=\mu^2) \quad.
\label{eq:thirtynine}
 \end{equation}
Combining the above results, we obtain at the final expression:
 \begin{equation}
M_B = M_B^{Born} \ e^{-\frac{\alpha_s}{4\pi} \, C_F  \ln^2(s/Q^2)} \quad .
\label{eq:forty}
 \end{equation}

\section {Non-forward structure functions}

The  DVCS structure functions are a generalization of the
conventional DIS structure functions.
 The novel feature  DVCS structure functions
is the explicit dependence on the fraction $\zeta$ of the longitudinal
momentum $r$ transferred to a hadron, $r=\zeta p$. The $Q^2$ evolution
of these structure functions is treated in the LLA in the framework
of a DGLAP type of equation\cite{Rad}.
Here we consider the non-forward quark structure function
in the DLA,
which we have applied above to the forward and backward DVCS
scattering amplitudes.

We can factorize the amplitude of the DVCS process and express the
quark structure function as a four-quark amplitude
$M_{ij,sp}(p,k,k-r)$, ({\it cf.}, Fig.5).
The two lower quarks in the amplitude $M_{ij,sp}$ carrying the momenta
$p$ and $p-r$ are on the mass shell,  while two upper ones with the
momenta $k$ and $k-r$ are virtual. The spinor indices $\{i,j\}$ belong
to the lower lines, and they are summed with the indices of the initial
and final quark spinors $U_j(p)$  and $\overline{U}_i(p-r)$. 
 The total amplitude is given by the integral of the function:
$$
\overline M_{sp}(p,k,k-r)
\ =\
\overline{U}_i(p-r) \
M_{ij,sp}(p,k,k-r) \
U_j(p)
$$
with the elementary quark-photon scattering amplitude:
\begin{eqnarray}
T^{\mu \nu} &=&
-\int \frac{d^4k}{(2\pi)^4} \  {\rm Tr} \, \overline M(p,k,k-r) \
\times
\nonumber \\
&\times&
\left[ \frac{\hat k}{k^2} \ \gamma^\mu \
\frac{\hat k + \hat q}{(k+q)^2} \ \gamma^\nu \
\frac{\hat k - \hat r}{(k-r)^2} \nonumber
+\frac{\hat k}{k^2} \ \gamma^\nu \
\frac{\hat k - \hat q^\prime}{(k-q^\prime)^2} \ \gamma^\mu \
\frac{\hat k - \hat r}{(k-r)^2} \
\right] \quad ,
\end{eqnarray}
The trace in $T^{\mu \nu}$ is taken over the upper line indices $\{s,p\}$.

\xfigv  

To derive the IREE for the amplitude $M_{ij,sp}(p,k,k-r)$, we
decompose this using a complete set of 16 gamma matrices:
$$
M_{ij,sp} \ = \ \sum_{A,B=1}^{16} \  \gamma^A_{si} \  \gamma^B_{jp} \ M_{AB}.
$$
where the sum runs over the basis set
$\gamma^i=\bigl\{1,\gamma^\lambda, \sigma^{\lambda \sigma},
\gamma^\lambda \gamma^5, \gamma^5 \bigr\}$.

In the double logarithmic approximation (DLA),
the scalar function $M_{AB}$ is comprised of
a product of the Born terms which can depend on the invariant energy
$s = (k-p)^2$ and the virtualities $l_1^2 = k^2$, $l_2^2 = (k-r)^2$.
Note the  $u=(k+p-r)^2$ invariant is not independent for $t\approx 0$ since
$s+u=l_1^2+l_2^2$.

The essential contribution of the amplitude $M_{ij,sp}$ comes only in
the region where both virtualities $\{l_1^2,l_2^2\}$
are negative and of the same order, {\it i.e.},
$l_1^2 \sim l_2^2 \sim l^2=-l^2_\perp$.
Therefore, it is natural to write  $M_{ij,sp}$  as a sum of an
s-channel and a u-channel piece:
\begin{eqnarray}
M_{ij,sp} \ &=& \ \frac{1}{s} \ \sum_{A,B} \
(\gamma^A)_{si}\  (\gamma^B)_{jp} \ F^S_{AB}
\left(\frac{s}{\mu^2},\frac{l^2}{\mu^2} \right)
\nonumber\\
&+& \
\frac{1}{u} \  \sum_{A,B} (\gamma^A )_{pi}
(\gamma^B )_{js} \ F^U_{AB}
\left(\frac{-u}{\mu^2},\frac{l^2}{\mu^2}\right).
\nonumber
\end{eqnarray}

It is easy to verify that there is no double logarithmic (DL)
contribution to the IREE in the
kinematic region $l^2 \not=0$ since $l^2_\perp$ plays the role
of the infrared cut-off (instead of $\mu^2$).
Thus,
\begin{eqnarray}
\left(s\frac\partial{\partial s} + \frac\partial{\partial z} \right) \
F^S_{AB}  &=& 0 \quad,
\nonumber \\
\left(u\frac\partial{\partial u} + \frac\partial{\partial z} \right) \
F^U_{AB}  &=& 0 \quad,
\end{eqnarray}
where we define
$$
z \ = \ \ln\frac{l^2}{\mu^2}  \quad .
$$
In terms of a Mellin transform,
$$
F^S_{AB} \ = \ \int\frac{d\omega}{2\pi i}
\left(\frac s{\mu^2}\right)^\omega \ f^S_{AB}(\omega,z)
$$
with a corresponding definition for $u$-channel function $F^U_{AB}$.
These functions satisfy the differential equation:
$$
\frac\partial{\partial z} \ f^{S,U}_{AB} \ = \ -\omega \ f^{S,U}_{AB} \quad .
$$
We can write the solution as:
$$
f^{S,U}_{AB}(\omega,z) \ = \ R^{S,U}_{AB}(\omega) \ e^{-\omega z} \quad ,
$$
where the function $R^{S,U}_{AB}$ is determined by the boundary conditions in the
limit where virtualities become
vanishingly small, $l^2 \sim \mu^2$ which implies $z=0$.
This kinematic region is the Regge limit which describes
the scattering of on-shell quarks with approximately zero momentum transfer.

The symbol $R^{S,U}_{AB}$ is the coefficient of
the Regge amplitude when expanded in the  $\{\gamma^A,\gamma^B\}$ basis.
 We can introduce an amplitude $A_{\rm Regge}^{h,h^\prime,f,f^\prime}$
which corresponds to the amplitude  $M_{ij,sp}$ in the Regge limit, and
expand this in the  $\{\gamma^A,\gamma^B\}$ basis as follows:
\begin{eqnarray}
A_{\rm Regge}^{h,h^\prime,f,f^\prime} \ &=& \
\sum_{ijsp}\overline U^h_s(q^\prime)
U^{h^\prime}_i(p) \ T^{\rm Regge}_{ij,sp} \
\overline U^{f^\prime}_j(p) U^f_p(q^\prime)
\nonumber \\
&=& \ \frac 1s \ \sum_{A,B} \ \overline U^h(q^\prime)
\gamma^A U^{h^\prime}(p) \cdot
\overline U^{f^\prime}(p)\gamma^B U^f(q^\prime) \
R^S_{AB} \nonumber \\
&+& \ \frac 1u  \ \sum_{A,B} \ \overline U^f(q^\prime)
\gamma^A U^{h^\prime}(p)\cdot \overline U^{f^\prime}(p)
\gamma^B U^h(q^\prime) \ R^U_{AB} \quad , \nonumber
\end{eqnarray}
where $h$, $f$, $h^\prime$, and $f^\prime$ are the quark helicities.

The spinor structure of the Regge amplitude is  determined
by the Born terms, so we have
$$
T^{\rm Regge}_{ij,sp} \ = \ \frac 1s \ \gamma^\lambda_{si}
\gamma^\sigma_{jp} \ g^{\lambda \sigma} \ R_S \ + \
\frac 1u \ \gamma^\lambda_{pi} \gamma^\sigma_{js} \ R_U \quad .
$$
In the Regge theory framework, it is more natural to work with amplitudes
of definite signature rather than s-channel and u-channel amplitudes,
rather than with the forward and backward ones, therefore, we
work with $R^{\pm}$ defined by:
\begin{eqnarray}
R_S &=& \ R^{(+)} \ + \ R^{(-)}
\nonumber \\
R_U &=& \ R^{(+)} \ - \ R^{(-)} \quad .
\end{eqnarray}
The form of the Regge amplitude dictates the final structure for the
four-quark amplitude:
\begin{eqnarray}
M_{ij,sp}(s,l^2) \ &=& \ \frac 1s  \
\bigl(\gamma^\lambda)_{si}\bigl(\gamma^\sigma)_{jp} \ g^{\lambda \sigma}
\int \frac{d\omega}{2\pi i} \ R_S(\omega)
\left(\frac s{l^2}\right)^\omega
\nonumber\\
&+& \
\frac 1u \  \bigl(\gamma^\lambda)_{pi}\bigl(\gamma^\sigma)_{js} \
g^{\lambda \sigma}
\int \frac{d\omega}{2\pi i} \ R_U(\omega)
\left(\frac s{l^2}\right)^\omega \quad .\nonumber
\end{eqnarray}

To extract the non-forward structure function,
the four-quark amplitude must be substituted into the total amplitude.
 For $t \approx 0$, the total amplitude is given by the sum of the
two transverse tensors,
$$
T^{\mu \nu} \ = \ \frac 1{2p \cdot q^\prime} \ g^{\mu \nu}_\perp \ T_1 \ - \
\frac 1{2p \cdot q^\prime} \ i\varepsilon^{\mu \nu \rho \eta} \
\frac 1{p \cdot q^\prime} \ T_2  \quad ,
$$
where $T_1$ is symmetric, and $T_2$ is antisymmetric.
We can express  $T_1$ and $T_2$ in terms of
the quark ($q$) and antiquark ($\bar q$) non-forward structure functions
${\cal F}^{q,\bar q}_\zeta(x)$ and $G^{q,\bar q}_\zeta(x)$,
\begin{eqnarray}
T_1 &=& \overline U(p-r)\hat q^\prime U(p) \
\int_0^1 dx \ \biggl[\frac 1{x-\zeta+i\delta} \ + \ \frac 1{x-i\delta}
\biggr] \ \bigl({\cal F}^q_\zeta(x) + {\cal F}^{\bar q}_\zeta(x)\bigr)
\quad ,
\nonumber \\
T_2 &=& \overline U(p-r)\hat q^\prime \gamma^5 U(p) \
\int_0^1 dx \ \biggl[\frac 1{x-\zeta+i\delta} \ - \ \frac 1{x-i\delta}
\biggr] \ \bigl(G^q_\zeta(x) + G^{\bar q}_\zeta(x)\bigr)
\quad .
\end{eqnarray}
(Here we follow the definitions given in Refs.\cite{Rad}.)

Introducing the Sudakov variables,  the explicit expressions for
$T_{1,2}$ are:
\begin{eqnarray}
T_1 \ &=& \ \int \frac{d^4k}{(2\pi)^4 i} \ {\rm Tr} \ \overline M(p,k,k-r)
\hat k\hat q^\prime (\hat k - \hat r) \ \frac 1{k^2(k-r)^2}
\nonumber \\
&\times&
\left[\frac 1{\beta-\zeta+i\delta} \ + \
\frac 1{\beta-i\delta}\right] \quad ,\nonumber
\\
T_2 \ &=&
-\int \frac{d^4k}{(2\pi)^4 i} \ {\rm Tr} \ \overline M(p,k,k-r)
\hat k\hat q^\prime \gamma^5 (\hat k - \hat r) \ \frac 1{k^2(k-r)^2}
\nonumber \\
&\times&
\left[\frac 1{\beta-\zeta+i\delta} \ - \
\frac 1{\beta-i\delta}\right] \quad ,
\nonumber
\end{eqnarray}
with
\begin{eqnarray}
d^4k &=& \frac{s^\prime}2 \ d\alpha \  d\beta \ dk_\perp  \quad ,
\nonumber \\
s^\prime &=& 2p \cdot q^\prime  \quad.
\end{eqnarray}
The integrals over $\beta$ are restricted  to the interval
$0\le \beta \le 1$ which follows from the position of the poles
in the $\alpha$-plane.
 Examining these expressions demonstrate that to
find the structure functions,  the amplitude $\overline M$ must be integrated
over the virtual momentum $k$ while keeping  $\beta$ fixed.
 In performing this integration, we need only keep the terms of the form
$d\alpha/\alpha  \ d k_\perp^2/k_\perp^2$ for DL accuracy.
These terms come from the region where
\begin{eqnarray}
k^2 \ &=& \ \alpha \beta s^\prime \ - \ k_\perp^2 \ \approx  \ - \ k_\perp^2
\nonumber \\
(k-r)^2 \ &=& \ \alpha (\beta -\zeta)s^\prime  \ - \ k_\perp^2
 \ \approx  \ - \ k_\perp^2 \nonumber \\
s \ = \ (k-p)^2 \ &=& \ \alpha (\beta -1)s^\prime  \ - \ k_\perp^2
 \ \approx  \ \alpha (\beta -1)s^\prime \nonumber \\
u \ = \ (k+p-r)^2 \ &=& \ \alpha (\beta +1-\zeta)s^\prime  \ - \ k_\perp^2
 \ \approx  \ \alpha (\beta +1-\zeta)s^\prime. \nonumber
\end{eqnarray}

These relations simplify the process of extracting the Regge limit.
First, they imply that both virtualities $\{l_1^2,l_2^2\}$
in the four-quark amplitude must be of the same order, {\it i.e.},
$l_1^2\sim l_2^2 \sim -k_\perp^2$;
this  justifies the assumption that was adopted previously.
Second, only the
quadratic terms in $k_\perp$  from the spinor
trace are necessary for the non-forward amplitudes $T_{1,2}$.
We can then compute these traces as follows:
\begin{eqnarray}
{\rm Tr} \ \overline M \ \hat k_\perp \hat q^\prime \  \hat k_\perp
 \ &=& \  \overline U(p-r) \ \gamma^A \  \hat k_\perp \hat q^\prime
\hat k_\perp
\gamma^B \  U(p) \ \bigl(\frac 1s F^S_{AB} \ + \ \frac 1u F^U_{AB}\bigr)
\nonumber \\
 \ &=& \ -2 \bigl(\frac 1s R_S \ + \ \frac 1u R_U\bigr) \ k^2_\perp \
\overline U(p-r) \ \hat q^\prime \ U(p) \nonumber\\
{\rm Tr} \ \overline M \ \hat k_\perp \hat q^\prime \gamma^5 \  \hat k_\perp
 \ &=& \  \overline U(p-r) \ \gamma^A \  \hat k_\perp \hat q^\prime \gamma^5
\hat k_\perp
\gamma^B \  U(p) \ \bigl(\frac 1s F^S_{AB} \ + \ \frac 1u F^U_{AB}\bigr)
\nonumber \\
 \ &=& \ 2 \bigl(\frac 1s R_S \ + \ \frac 1u R_U\bigr) \ k^2_\perp \
\overline U(p-r) \ \hat q^\prime \gamma^5 \ U(p)
\quad .
\nonumber
\end{eqnarray}

To calculate the structure functions, we must replace
the infrared cut-off $\mu^2$ in the Regge amplitude by
$-k_\perp^2$, and then integrate over the loop momentum.
The forward and backward amplitudes will yield the
quark and antiquark functions, respectively.
In the DL region,  the transverse momentum is
$$
k_\perp^2 \ \ge \ Max
\left\{|\alpha \beta |s^\prime  \ , \
|\alpha (\beta - \zeta) |s^\prime \right\}
\quad .
$$
The remaining integral over  $\alpha$  is purely logarithmic,
and the symmetric integration limit yield  $ i \pi$.

Finally, working with DL accuracy we retain only the
non-vanishing terms in the limit $\omega \to 0$ to arrive at the following
structure functions:
\begin{eqnarray}
{\cal F}^q_\zeta(\beta) \ &=& \ \frac 1{8 \pi^2} \ C_F \
\frac 1{1-\beta} \ \int \frac{d\omega}{2\pi} \ R_S(\omega) \ \frac 1\omega \
\times
\nonumber \\
&\times&
\left[\theta(\beta-\frac 12 \zeta) \ \left(\frac{1-\beta}\beta \right)^\omega
+ \theta(\frac 12 \zeta-\beta) \ \left(\frac{1-\beta}{\zeta -\beta}
\right)^\omega \right],
\label{eq:fquark}
\end{eqnarray}
\begin{eqnarray}
{\cal F}^{\bar q}_\zeta(\beta) \ &=& \ \frac 1{8 \pi^2} \
C_F \ \frac 1{1+\beta-\zeta} \
\int \frac{d\omega}{2\pi} \ R_U(\omega) \ \frac 1\omega \
\times
\nonumber \\
&\times&
\left[\theta(\beta-\frac 12 \zeta) \ \left(\frac{1+\beta-\zeta}\beta
\right)^\omega
+ \theta(\frac 12 \zeta-\beta) \ \left(\frac{1+\beta-\zeta}
{\zeta -\beta}\right)^\omega \right]
\quad .
\label{eq:fqbar}
\end{eqnarray}

The explicit expressions for $R_S$, $R_U$ can be
taken from Refs.~\cite{KL,BFKL}.
The functions $G^{q, \bar q}_\zeta(\beta)$ are expressed by
Regge amplitudes using the same formula. The integral over $\omega$
runs along imaginary axes to the right of the singularities of the
$R_{S,U}$ functions.
Non-zero contributions exist only when
the arguments in the brackets are larger than unity, so that the
actual support
of the first term on the RHS of \eq{fquark} and \eq{fqbar}
is $\zeta/2 \le \beta \le 1/2$.
The Regge limit implies the argument of the Mellin transform is
be asymptotically large, which corresponds to the the region where
$\zeta \ll 1$ and $\beta \sim \zeta$.

The absence of a $Q^2$ dependence in these expressions resembles
the DGLAP type of evolution equation for the non-forward
structure functions in $Q^2 \to \infty$ limit.
In contrast to the DGLAP evolution equations, the DVCS
evolution equations are not restricted to the region $\beta \le \zeta$.

Notice that for $\beta=0$, we find that
${\cal F}^{q,\bar q}_\zeta(0) \not = 0$ in contrast to the
result for the DGLAP non-forward functions.
The Mellin moments
$M_n^{q,\hat q}(\zeta) = \int_0^1 dx \, x^{n-1} {\cal F}^{q,\hat q}_\zeta(x)$
for the integer $n$ are not necessarily polynomials of
degree $n$ with respect to the transfer momentum fraction $\zeta$,  as in
the DGLAP case.

\section{Conclusion}

We obtained the explicit expressions for the DVCS amplitudes
in the Regge limits of forward and backward scattering.
For the forward scattering region, the asymptotic amplitude
takes the form of the DIS structure functions times a Sudakov
exponential factor of the form
$exp[-(\alpha_s/4\pi) \, C_F \ln^2(-t/\mu^2)]$.
 The DVCS amplitude in the backward scattering region is  purely
of the Sudakov type with an
exponential factor of the form
$exp[-(\alpha_s/4\pi) \, C_F \ln^2(s/\mu^2)]$.
 We have not used the DGLAP evolution equations which
assume the transverse momenta to be strictly ordered in
virtuality. This enables us to take into account important
logarithmic contributions that are beyond of the realm of
the DGLAP approach.
 Specifically, there is no ordering of the transverse momenta
in the Regge region, and this allows the internal transverse momenta
to exceed  $Q^2$, in contrast to the DGLAP method.
Such contributions lead to the power-like behavior of the DVCS amplitude.

\section{Acknowledgments}

 This work is supported in part by
 Grant INTAS-RFBR-95-311,
 Grant RFBR-98-02-17629,
 the US Department of Energy, and the 
Lightner-Sams Foundation. 
 BIE thanks the SMU Theory Group for their kind
hospitality and the Lightner-Sams Foundation for
financial support  during the  period in which part
of this research was carried out.


\end{document}